\documentclass[]{aa}

\usepackage{graphicx,times}
\usepackage{graphics}

\begin{document}

   \title{XMM-Newton observations of the Cluster of Galaxies Abell~496}

   \subtitle{Measurements of the elemental abundances in the intracluster medium}

   \author{
	T. Tamura \inst{1}
	\and
	J. A. M. Bleeker \inst{1}
          \and
	J. S. Kaastra \inst{1}
	\and
         C. Ferrigno \inst{1}
         \and
	S. Molendi \inst{2}	
	}

	   \offprints{T.Tamura}
	\mail{T.Tamura@sron.nl}

\institute{ SRON National Institute for Space Research, 
              Sorbonnelaan 2, 3584 CA Utrecht, The Nether\-lands 
	\and
	IFC Milan, 20133 Milano, Italy
}

   \date{Received 2001; accepted 13 September 2001}

\abstract{
The results from {\it XMM-Newton} observations of the relaxed cluster of galaxies Abell~496 are presented.
The spatially-resolved X-ray spectra taken by the European Photon Imaging Cameras show a temperature drop 
and a Fe abundance increase in the intra-cluster medium (ICM) towards the cD galaxy at the cluster center.
The abundances of Si and S also show a central enhancement.
High resolution soft X-ray spectra obtained with the Reflection Grating Spectrometers 
provides a strong constraint on the temperature structure in the central cool plasma.
Furthermore, the O abundance at the cluster core is accurately measured based on the \ion{O}{viii}~${\rm Ly\alpha}$ line detected with the RGS.
Contrary to the Si, S, and Fe abundances, the O abundance is radially constant over the cluster.
}

   \maketitle
%

\section{Introduction}
Clusters of galaxies are filled with an X-ray emitting ICM.
Measurements of X-ray lines from ionized heavy elements such as Fe have indicated that the ICM contains a large amount of metals.
These metals have been supposedly produced by supernovae (SNe) in member galaxies and ejected into the intergalactic space.
Therefore, the distribution of heavy elements in the ICM is essential for understanding the evolution of clusters.
Following early measurements, 
{\it ASCA} and {\it BeppoSAX} observations have revealed several important properties of the ICM metallicity.
These include Fe abundance increases around cD galaxies (e.g. Fukazawa \cite{fukazawa-phd}; De Grandi and Molendi \cite{grandi}) 
and variations in Si/Fe ratio within a cluster (e.g. Finoguenov et al. \cite{finoguenov00}) and among clusters (e.g. Fukazawa et al. \cite{fukazawa98}).
The variations in Si/Fe suggest that the metals in the ICM have been produced by at least two different origins, 
presumely in SN Ia and in SN II.

To further understand the metal enrichment in the ICM, 
accurate measurements of the distribution of heavy elements are required.
In particular, the O abundance (or O/Fe ratio) is important and more sensitive to its origin 
since O is expected to be produced mostly by SNe II.
Nevertheless, accurate abundances of O have not been measured, 
except for a few cases (e.g. Matsumoto et al. \cite{matsumoto}), 
due to limited sensitivities of pre-{\it XMM-Newton} instruments for the \ion{O}{viii} lines.

Here we report the first results from the {\it XMM-Newton}  (Jansen et al. \cite{jansen}) observations of A~496.
This cluster, at a redshift of 0.033, 
was studied in detail with previous X-ray instruments, 
including $Einstein$ (e.g. Nulsen et al.~\cite{nulsen}), 
{\it ASCA} (e.g. Markevitch et al.~\cite{markevitch}; Dupke and White 2000, \cite{dupke} hereafter).
These studies have shown that the ICM has a temperature distribution of 2--5~keV without any evidence for a major merger, 
indicating that this is one of the brightest clusters suitable for measurements of elemental abundances.
Using spatially-resolved spectra, 
we have measured elemental abundances including O
and their spatial distributions in the ICM within 6\arcmin.4 in radius.
We found clear Fe-, Si-, and S- concentrations around the cD galaxy
and a change in O/Fe over the cluster.

Throughout this paper,
we assume the Hubble constant to be $H_0 = 100h$ km s$^{-1}$Mpc$^{-1}$
and use the 90\% confidence level unless stated otherwise.
One arc-minute corresponds to $28h^{-1}$~kpc.

\section{Observations}
{\it XMM-Newton} observations of A~496 were performed on 2000 March 13--15 and 2001 February 1.
For the EPIC analysis, 
we used the on-axis and $\sim10$\arcmin off-axis observations.
The three EPIC cameras (MOSs and PN; Turner et al. \cite{turner}; Str\"{u}der et al. \cite{struder})  
were operated in the full window mode with the thin filter.
After removing high background periods, 
we obtained useful exposure times of 10--15~ksec and 34--35~ksec for the on-axis and off-axis pointings, respectively.
For the RGS (den Herder et al. ~\cite{herder}) analysis, 
we used the on-axis observation 
with a useful exposure of 21~ksec.
The RGS dispersion axis is oriented along an axis of about $-100$ degrees (North to East).

   \begin{figure}
\resizebox{\hsize}{!}{\includegraphics[]{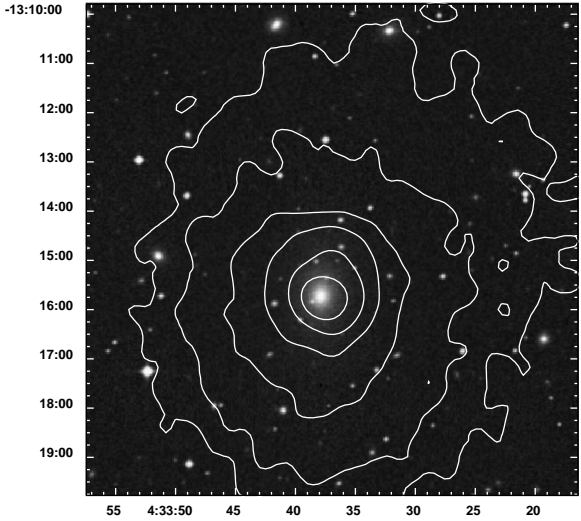}}

	      \caption[]{X-ray contour image of A~496 taken with the {\it XMM-Newton} MOS1 in the 0.9--1.2 keV band.
	A $10' \times 10'$ optical image obtained from the STScI Digitized Sky Survey provided by Leicester Database and Archive Service is overlaid.
	The X and Y axis are Right Ascension and Declination in J2000 coordinates, respectively.
	The X-ray image has been smoothed with a Gaussian filter with $\sigma=12$\arcsec.8.
	No correction for the vignetting or background was made on the X-ray image.
	The contours are separated by factors of two.
	}
         \label{fig:image}
   \end{figure}

\section{The EPIC results}
Figure~\ref{fig:image} shows an X-ray surface brightness of A~496 in the 0.9--1.2 keV band (around the Fe-L complex) with the MOS1 image overlaid on the optical image. 
The X-ray emission is thus distributed azimuthally symmetrically centered on a cD galaxy, MCG-02-12-039.

For basic data processing, we used the Science Analysis System (SAS).
The data were corrected for telescope vignetting.
For spectral fitting, we utilize the SPEX package (Kaastra et al. \cite{kaastra}).
We model the plasma emission using the MEKAL emission code (Mewe et al. \cite{mewe}).
The solar abundances are taken from Anders and Grevesse (\cite{anders}).

The background was taken from Lockman-Hole observations and subtracted before the spectral fitting.
The background uncertainty could introduce systematic error on the fitting results 
for the analysis of clusters with {\it XMM-Newton}.
The EPIC background is dominated by the cosmic diffuse emission at low energies.
This component varies significantly with the sky position.
For example, Ishisaki (\cite{ishisaki}) found about 50\% variation in the soft X-ray flux of the cosmic background based on the {\it ASCA} observations of several blank fields. 
At high energies, the EPIC background is dominated by
a particle-induced component.
In order to estimate the variation in this component, 
we examined several blank-sky field observations of the MOS and found the variation to be less than 20\%.
Accordingly, 
we assigned 50\% and 20\% systematic errors to the background below and above 1.2~keV,
respectively.
In addition, 3\% systematic errors to the source count are introduced to represent uncertainties in both the plasma model and instrumental response. 
We took both these systematic errors and the statistical errors
into account when assessing the best-fit parameters and their errors.
To avoid additional calibration uncertainties of the response and background subtraction, 
we used only the 0.4--8.0~keV and 0.5--8.0~keV energy bands for MOS and PN, respectively.

\subsection{Radial distribution of the ICM properties}
We extracted the spectra in several annuli around the emission center with outer radii ranging from $16''$ to $512''$.
We fitted the MOS1, MOS2, and PN spectra for on- and off-axis exposures separately 
with a single temperature plasma model modified by photoelectric absorption.
Solar abundance ratios for the heavy elements were assumed.

\begin{figure}
	\resizebox{\hsize}{!}{\includegraphics[]{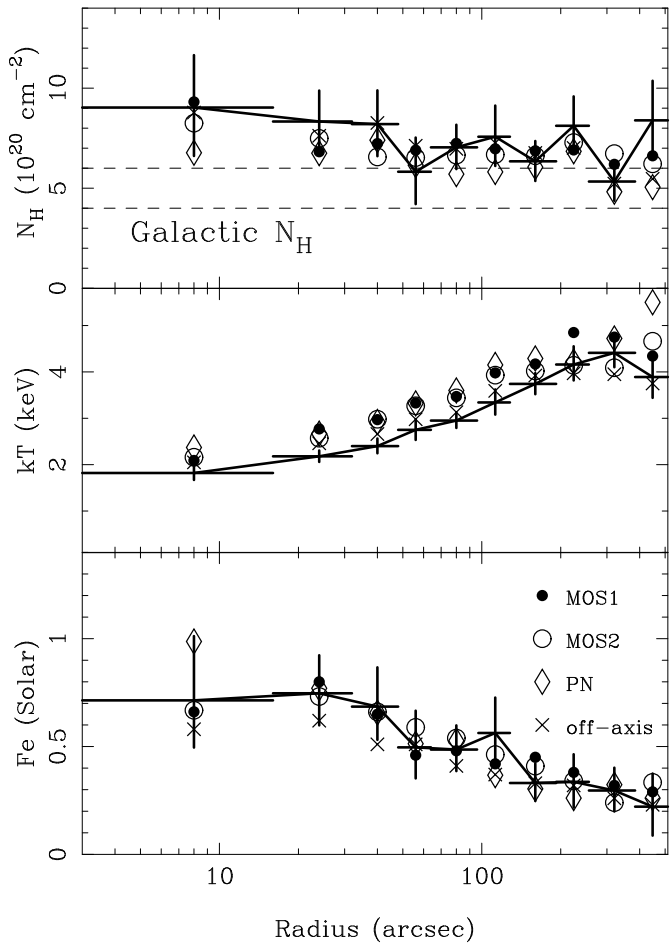}}
      \caption[]{Radial properties of the ICM 
       derived from the EPIC spectra based on a single temperature model. 
       From top to bottom, absorbing column density, temperature, 
	and Fe abundance are shown, respectively. 
	Filled circle, open circle, and diamond marks show the best-fit parameters from {\em projected} 
	 spectra of MOS1, MOS2 and PN, respectively, of the on-axis exposures.
	The marks '$\times$' show the average parameters from off-axis exposures.
	Bold lines with error bars show the results from simultaneous fitting of {\em de-projected} spectra.
	Errors in other results are similar to or smaller than the deprojected results.
	In the top panel, the range of the Galactic column density is indicated by dashed-lines.
}
         \label{fig:r-ta}
   \end{figure}

A single temperature model provided an adequate description of all the spectra.
The absorption column densities, temperatures, and metallicities are shown in Fig.~\ref{fig:r-ta}.
The metallicities were primarily determined by Fe-L and Fe-K line emissions.
The results from all instruments, for both on-axis and off-axis pointings, are consistent with each other within the errors.

In order to estimate the role of projection effects, 
we next created deprojected spectra using Plummer's method (e.g. Kaastra \cite{kaastra89}).
Here we assumed spherical symmetry of the emission distribution and no emission beyond the field of view ($\sim 15'$). 
These spectra were fitted to the same single temperature model.
Here and hereafter, we fitted the combined MOS (MOS1+MOS2) and PN spectra simultaneously.
All spectra can be described well by the single temperature model; 
$\chi^2$ ranges from 586 to 451 for $\nu = 531$.
In other words, the ICM can be well approximated by a single-phase spectral model for each radius.
As shown in Fig.~\ref{fig:r-ta}, the best-fit parameters are reasonably 
similar to those of the projected spectra obtained above, 
with slight drops of temperatures in central annuli.

The obtained column densities are marginally consistent with the Galactic value of 
$(4-6)\times10^{20}$~cm$^{-2}$ obtained from the \ion{H}{i} map (Dickey \& Lockman \cite{dickey} using NASA's W3nH tool).
There is no indication of excess absorption towards the cluster center.
Beyond a radius of 2\arcmin--3\arcmin\ from the center, 
the temperature continuously decreases towards the center from 4--5 keV down to 2 keV,
with a logarithmic slope of 0.25--0.3, while 
the iron metallicity increase towards the center between radii of 0.5\arcmin\ and 3\arcmin. 

We do not take into account the telescope PSF, which has a half energy width of $15$\arcsec.
Therefore the true profiles of temperature and metallicity at the cluster center could be much steeper than the present results.
On the other hand, results from outer regions, where the radial bin size is larger than the PSF width, should not change significantly.

   \begin{figure}
	\resizebox{0.90\hsize}{!}{\includegraphics[angle=-90]{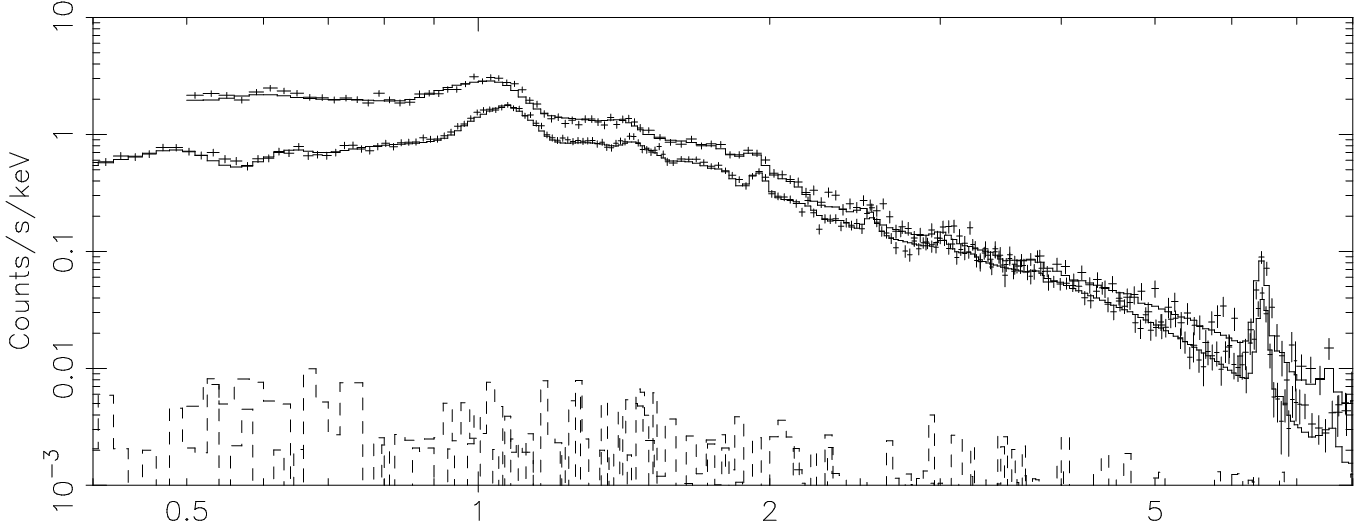}}
	\resizebox{0.90\hsize}{!}{\includegraphics[angle=-90]{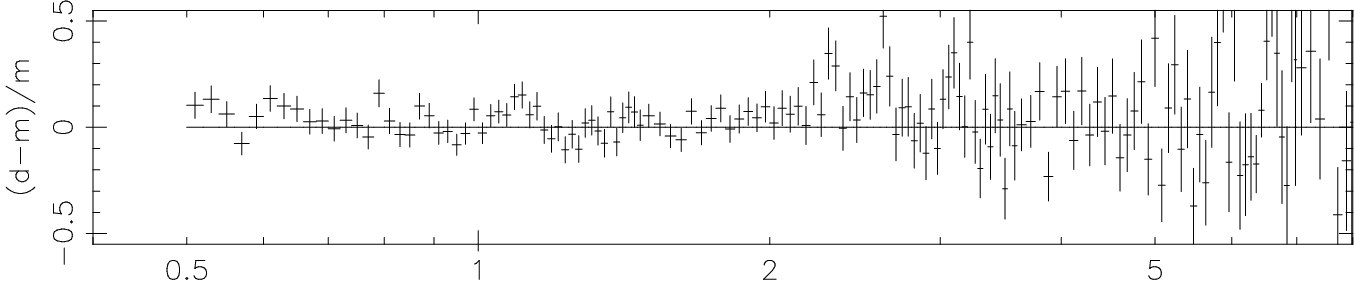}}
	\resizebox{0.90\hsize}{!}{\includegraphics[angle=-90]{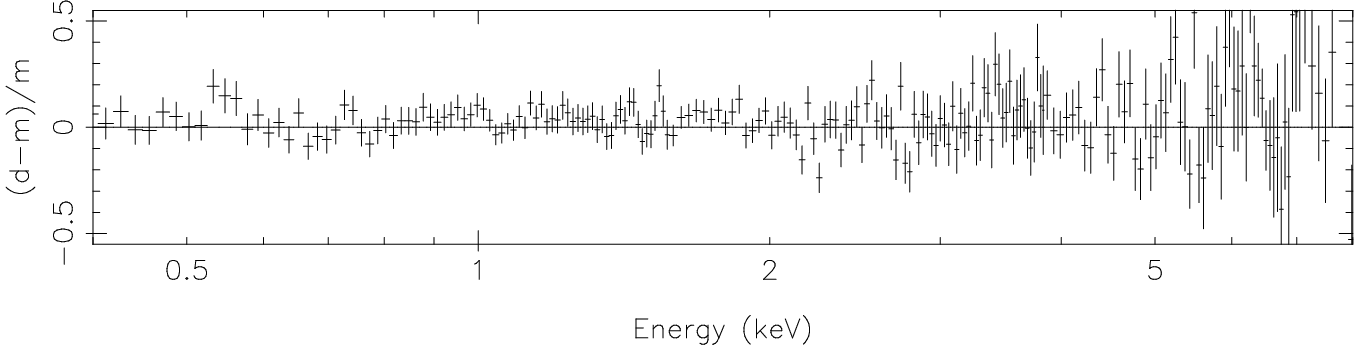}}
      \caption[]{Top panel: The EPIC PN (upper one) and MOS (lower) spectra extracted from 0\arcsec--48\arcsec\ in radius from the cluster center. 
The best-fit two temperature models with abundances of Fig.~\ref{fig:r-abun} and the background spectra are shown with full-line and dashed-line histograms, respectively.
Middle and bottom panels: The fit residuals [(data--model)/model] of the PN and MOS spectra, respectively}
         \label{fig:ring12}
	\resizebox{0.90\hsize}{!}{\includegraphics[angle=-90]{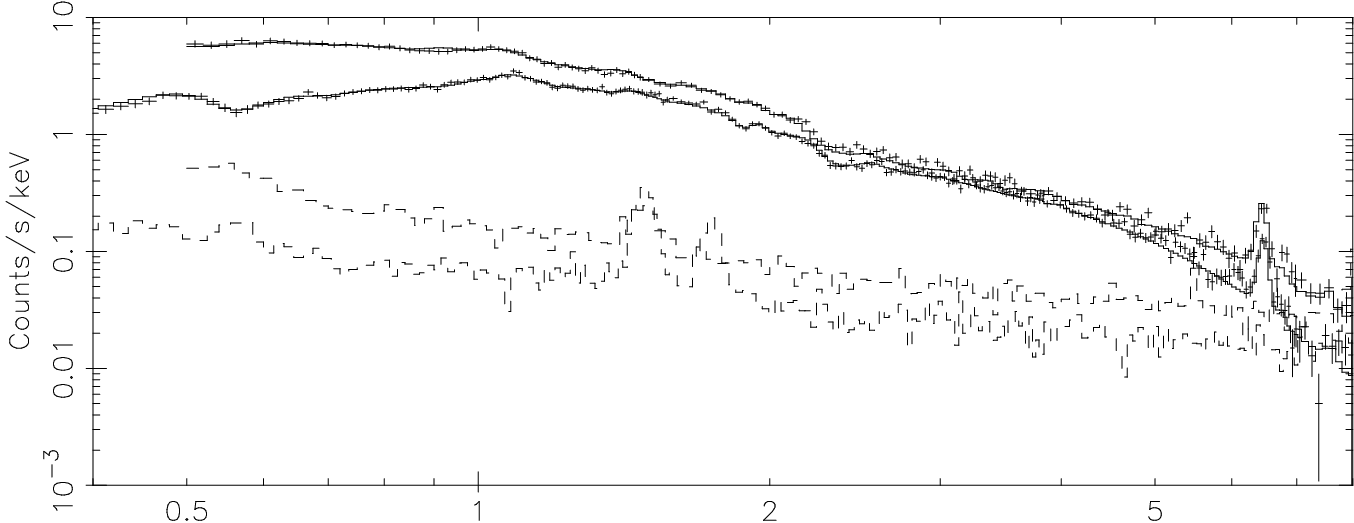}}
	\resizebox{0.90\hsize}{!}{\includegraphics[angle=-90]{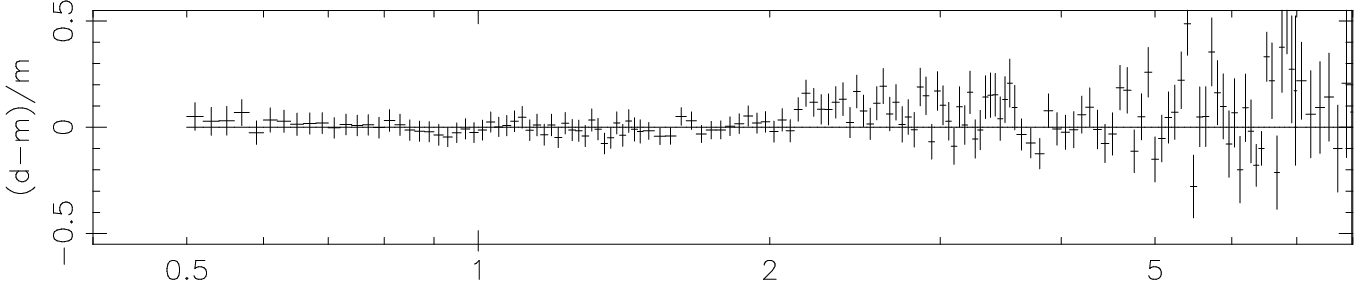}}
	\resizebox{0.90\hsize}{!}{\includegraphics[angle=-90]{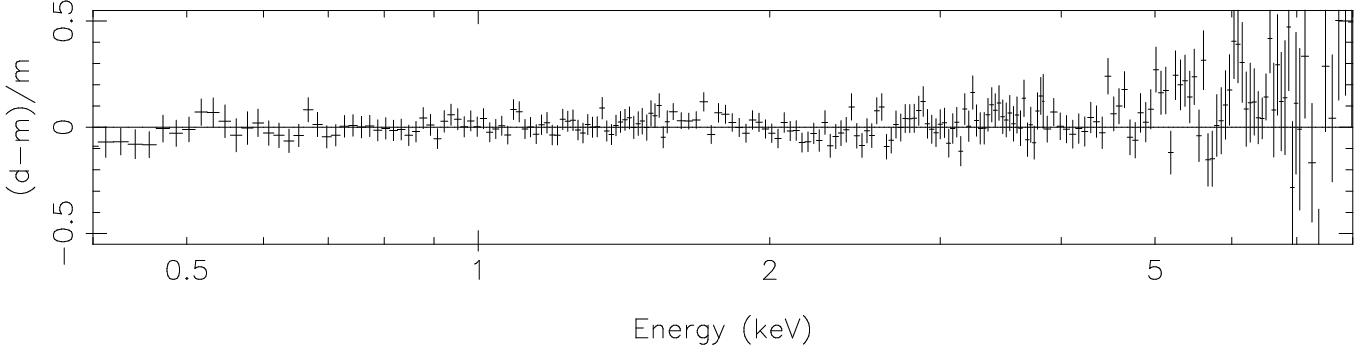}}

      \caption[]{Same as the previous figure, but spectra from the outer region (128\arcsec--384\arcsec\ in radius) with the best-fit single temperature model.}
         \label{fig:ring14}
   \end{figure}

\subsection{Elemental abundances} \label{sect:abun}
The relatively low temperature of A~496 and good sensitivity of the {\it XMM-Newton} instruments
provide an unique opportunity to determine the elemental abundances of the ICM
based on the K-line emission from O, Ne, Mg, Si, S, Ar, and Ca and K- and L-line emission from Ni and Fe.

In order to measure elemental abundances and their distribution with reasonable accuracy,
we divided the cluster emission into three regions, 
0\arcsec--48\arcsec, 48\arcsec--128\arcsec\, and 128\arcsec--384\arcsec, in radii.
These were chosen so that each region has enough photons and no significant variation in temperature.
To describe the temperature structure obtained above (Fig.~\ref{fig:r-ta}), 
the central spectrum (0\arcsec--48\arcsec) were fitted with a two temperature (2T) model.
On the other hand, the other two sets of spectra were fitted with a single temperature (1T) model.
The abundances of O, Ne, Mg, Si, S, Ar, Ca, Fe, and Ni were left free.
The abundances of C and N were coupled to the Fe value.
In the case of 2T model, the abundances of the two  components were set equal.

\begin{figure*}
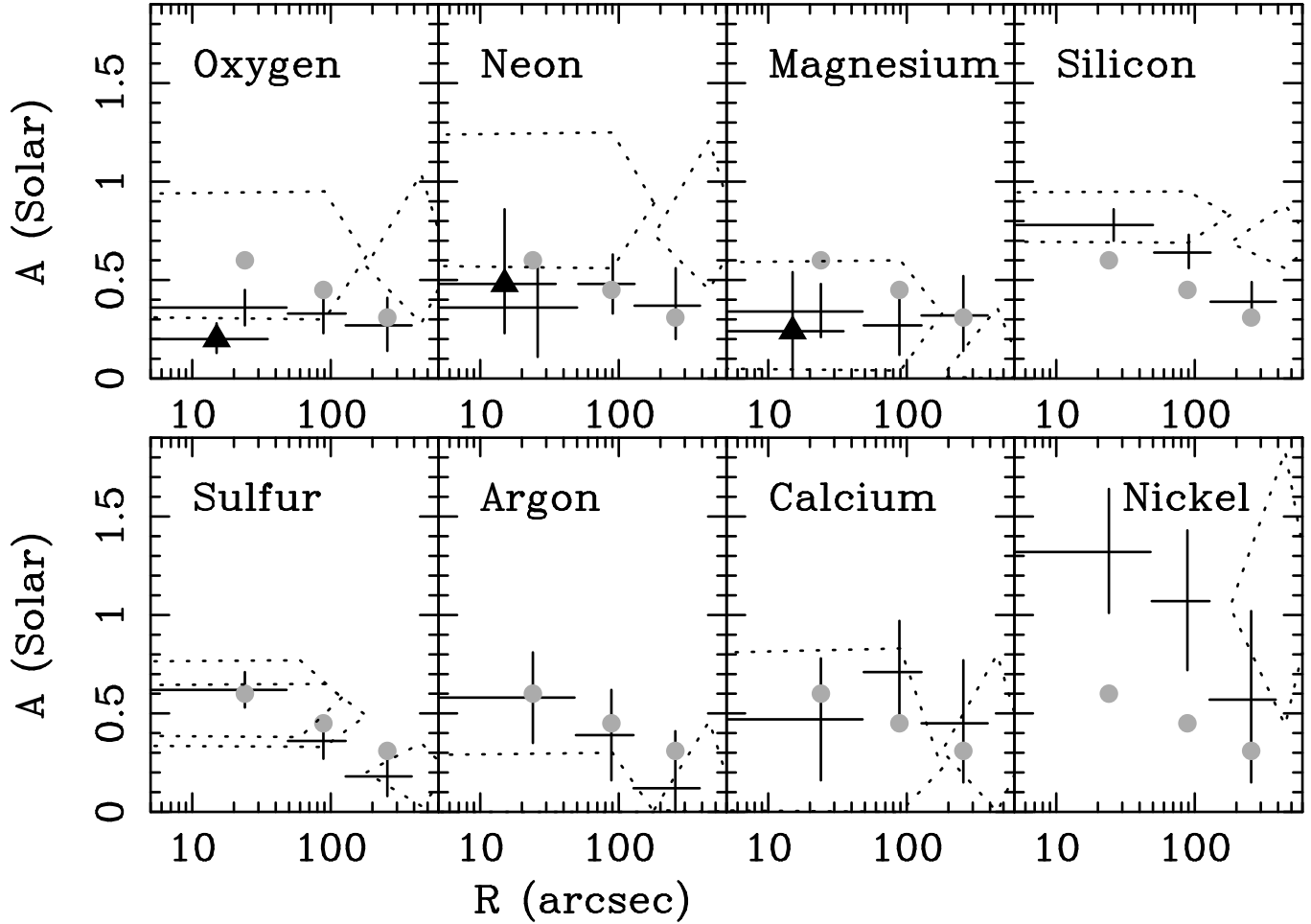

\begin{center}
	\resizebox{1.0\hsize}{!}{\includegraphics[angle=-90]{0710-1.qdp.ps}}
	\resizebox{1.0\hsize}{!}{\includegraphics[angle=-90]{0710-2.qdp.ps}}
\end{center}
      \caption[]{Radial distribution of elemental abundances with 90\% confidence limits. 
	The EPIC and RGS results are shown as crosses and crosses with triangle marks, respectively.
	For reference, the Fe abundance from EPIC is shown in each panel in gray-color-filled-circles. The errors of Fe abundance are less than 0.03 solar.
	The {\it ASCA} results (\cite{dupke}) are also shown by dashed-line-diamonds.
}
         \label{fig:r-abun}
   \end{figure*}

The obtained temperatures are consistent with the above results (Fig.~\ref{fig:r-ta}).
Examples of the fitting results are shown in Figs.~\ref{fig:ring12}-\ref{fig:ring14}.
We illustrate the derived abundances in Fig.~\ref{fig:r-abun}.
In addition to the Fe abundance, the Si and S abundances show a significant ($>4\sigma$ confidence) central enhancement.
The Ar and Ni abundance marginally ($\sim 2\sigma$ confidence) show increases.
Contrary to these elements, the O, Ne, and Mg abundances are close to being uniform.

\section{The RGS results}
\subsection{Isothermal model fitting}
The RGS first and second order spectra from RGS1 and RGS2 were extracted as follows.
The events were screened by rejecting the high background periods.
These events were filtered using both the dispersion vs. cross-dispersion and dispersion vs. pulse height windows.
The size of the spatial extraction region is 1\arcmin.1 in the cross-dispersion direction.

\begin{figure*}
	\resizebox{\hsize}{!}{\includegraphics[angle=-90]{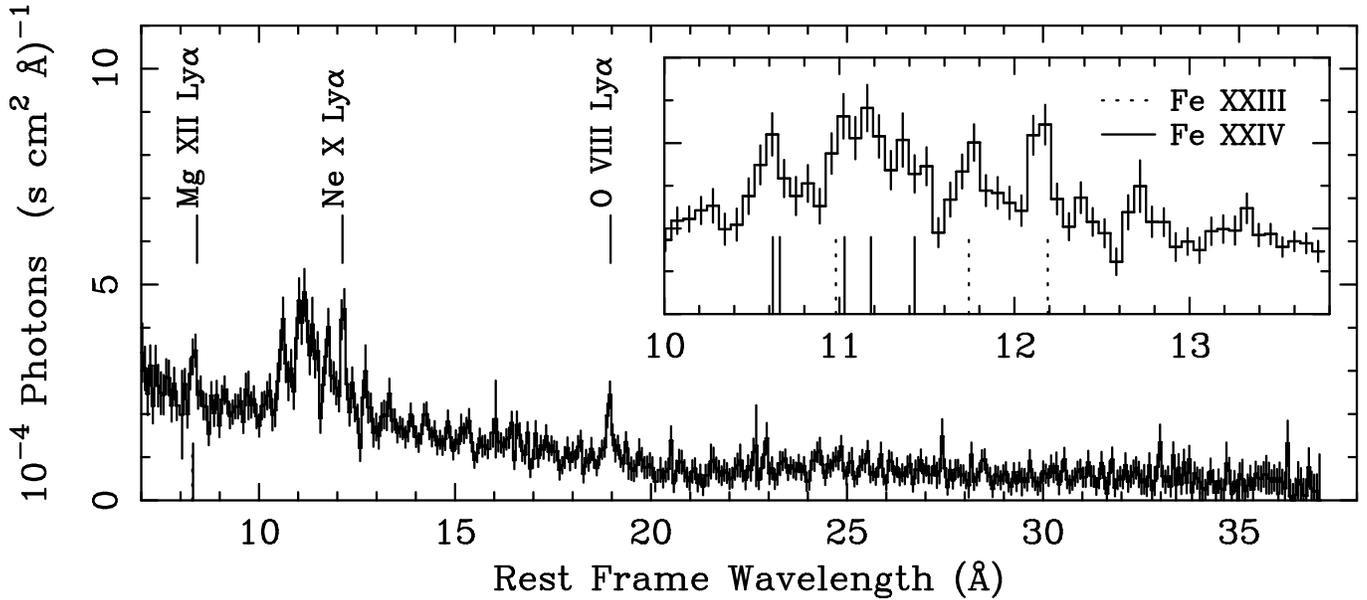}}
      \caption[]{The RGS spectrum extracted from the central 1\arcmin.1 in full-width. 
	The spectrum is corrected for effective area and redshift, but not corrected for the background.
	Several line positions are indicated;
	Mg XII (8.42\AA),  Ne X (12.13\AA), OVIII (18.97\AA), 
	Fe XXIII (8.30, 10.98, 11.74, 12.19\AA),  
	Fe XXIV (8.32, 10.62, 10.66, 11.03, 11.18, 11.43\AA).
	Insert is a zoom of a part of the spectrum.}
         \label{fig:rgs-spe}
   \end{figure*}

Figure~\ref{fig:rgs-spe} shows a combined spectrum which is corrected for the effective area.
Significant line emission from \ion{Mg}{xii}, \ion{Fe}{xxiii}, \ion{Fe}{xxiv} and \ion{Ne}{x}
as well as a clean line from \ion{O}{viii} have been detected.

The RGS energy responses were generated with the SAS tool (rgsrmfgen v.0.34).
An instrumental absorption feature close to the neutral O K edge (den Herder et al. ~\cite{herder}) 
was approximated by a photoelectric absorption due to \ion{O}{i} with a column density of $3\times10^{21}$~m$^{-2}$.
We calculated the line spread function based upon the source surface brightness profile
and convolved this with the RGS response for a point source.
A $\beta$ model was assumed for the brightness profile.
The core radius was determined to be $0'.45$ in radius by fitting the observed line shape of the \ion{O}{viii} in the RGS spectrum.
The Background spectrum and its variability were estimated from several blank-sky field observations.
We assigned 3\% and 30\% systematic errors to the source and the background counts, respectively.
We limited the wavelength band to 8--28~\AA\ and 8--14~\AA\ for the first and second order spectra, respectively,
where the estimated background is less than 20\% of the source flux.
The four spectra were fitted with the same model simultaneously, but with different normalizations.

We fitted the RGS spectra with an isothermal model as in the EPIC analysis.
The O, Ne, Mg, and Fe abundances were left free, while all the other elements were coupled to the Fe value.
The fits were statistically acceptable; 
the results are shown in Table~\ref{tbl:rgs-fits} and Fig.~\ref{fig:rgs-spe-fit}.
Being consistent with the EPIC results, there is no indication of an excess absorption.

\subsection{Limits on the central cool emission} \label{sect:cool}
\begin{figure}
	\resizebox{\hsize}{!}{\includegraphics[angle=-90]{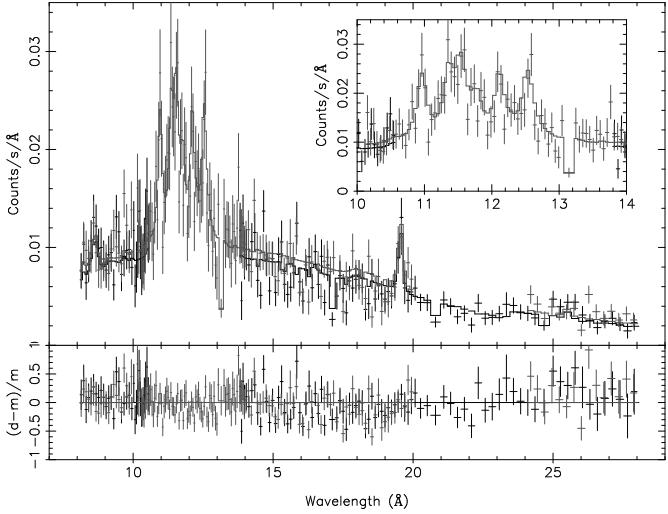}}
      \caption[]{The RGS first order spectra with the best-fit isothermal model (RGS1:black and RGS2:gray). The bottom panel shows the fit residuals.}
         \label{fig:rgs-spe-fit}
   \end{figure}

Thus, the RGS spectra can be described by the single temperature model
and show no indication of emission lines expected from a much cooler component (Fig.~\ref{fig:rgs-spe-fit}).
In addition, the temperature is close to that found with the EPIC in the cluster center.
Nevertheless, a much cooler component due to either the radiative cooling of the ICM or inter-stellar medium of the cD galaxy 
could possibly exist around the galaxy.

In order to constrain such a possible central cool component,
we fitted the RGS spectra by adding a cooler plasma component to the isothermal model.
Firstly, we considered an additional isothermal component with a temperature of 1~keV (2T model).
The abundances (O, Ne, Mg, and Fe) of both component were fixed to each other.
This 2T model gave a better fit than the 1T model (Table~\ref{tbl:rgs-fits}).
The emission measure of the 1~keV component ($EM_{\rm cool}$) is less than 5\% of that of the hot component.

Secondly, we used an isobaric cooling flow model (Johnstone et al. \cite{johnstone})
in addition to the hot isothermal component (1T+ICF model).
The ICF model is characterized by lower and upper limit temperatures ($T_{\rm min}$ and $T_{\rm max}$) and its normalization (the mass deposition rate, $\dot{M}$).
The RGS spectrum is not sensitive to the changes in temperature 
if the temperature is larger than $\sim 3$~keV and hence to the abundance of the isothermal component.
Therefore we fixed some parameters in accordance with the EPIC results,
i.e the temperature of the isothermal component was fixed to $T_{\rm max}=4$~keV 
and the Fe abundance was fixed to be 0.6 times solar.
If we fix $T_{\rm min}$ to a very small value (0.01~keV), the normalization of the ICF component ($\dot{M}$) becomes zero.
When $T_{\rm min}$ is left free, the fit slightly improves as shown in Table~\ref{tbl:rgs-fits}.
The $\chi^2$ for this fit (310) is close to that for the 2T model (303).
Since the RGS spectrum is dominated by the hot isothermal component,
it is difficult to measure the values of the weaker additional parameters.
We derived the 90\% confidence upper limit of $\dot{M}$ for a given $T_{\rm min}$.
For $T_{\rm min}$ = [0.3, 1.0, 1.5]~keV, $\dot{M}$ is less than [10,40, 80]~$h^{-2}$~\hbox{M$_{\odot}$}yr$^{-1}$.

The results of these fittings show that the O abundance and O/Fe ratio
do not differ significantly between models.

We performed a similar analysis using the EPIC spectra extracted from the central region (r~$<48$\arcsec) 
and obtained results  consistent with the RGS analysis.
We fitted the EPIC spectra with the 1T+ICF model.
The quality of the fit ($\chi^2/\nu = 1232/934$) is similar to the 2T model (Sect.~\ref{sect:abun}).
The best-fit model gives $T_{\rm min}$ and $\dot{M}$ of 1.5~keV and 70$~h^{-2}$~\hbox{M$_{\odot}$}yr$^{-1}$, respectively, 
consistent with the above RGS result.
The elemental abundances are also fully consistent with those obtained with the 2T model (Fig.~\ref{fig:r-abun}), 
indicating that the derived elemental abundances are rather invariant to the adopted model temperature structure.

\begin{table}
\caption[]{The fit results of the RGS spectra of the center of A~496. }
\label{tbl:rgs-fits}
\begin{center}
\begin{tabular}{lccc}
\hline 
model			& 1T 				& 2T	& 1T+ICF\\
\hline
$EM_{\rm hot}^a$	& 1.3				& 1.1		& 0.7\\
$T$ (keV)		& 2.2 ($_{-0.3}^{+0.4}$)	& 2.6		& 4.0$^f$\\
O$^b$			& 0.23 ($_{-0.07}^{+0.08}$)	& 0.29		& 0.29\\
Ne$^b$			& 0.48 ($_{-0.25}^{+0.33}$)	& 0.62		& 0.53\\
Mg$^b$			& 0.24 ($<0.54$)		& 0.43		& 0.35\\
Fe$^b$			& 0.43 ($_{-0.11}^{+0.15}$)	& 0.68		& 0.6$^f$\\
$N_\mathrm{ H}^c$	& 10.3 ($\pm 1.5$) 		& 8.8		& 8.9\\
$EM_{\rm cool}^a$	or $\dot{M}^d$& -		& 0.04 ($<0.05$) & 50 \\
$T_{\rm cool}$ or $T_{\rm min}$& -			& 1.0$^f$ 		& 1.2\\	
$\chi ^2/\nu$	& 314/358				& 303/357	& 310/357\\
\hline
\end{tabular}
\end{center}
\begin{description}
\item[$^a$] The volume emission measure in units of $h^{-2}10^{72}$m$^{-3}$.
\item[$^b$] Metal abundances relative to the solar values (
O/H$=8.51\times10^{-4}$, Ne/H$=1.23\times10^{-4}$, Mg/H$=3.8\times10^{-5}$ and Fe/H$=4.68\times10^{-4}$).
\item[$^c$] The column density in units of $10^{20}$~cm$^{-2}$.
\item[$^d$] The mass deposition rate in units of $h^{-2}$~\hbox{M$_{\odot}$}yr$^{-1}$.
\item[$^f$] These are fixed parameters.
\end{description}
\end{table}

\section{Discussion}
\subsection{Summary of the results}
Based on the {\it XMM-Newton} observations of A~496, we have measured the radial properties of the ICM.
We have confirmed the presence of relatively cool and metal-rich plasma 
around the cluster center. 
The temperature is spatially resolved; it continuously decreases towards the center (Fig.~\ref{fig:r-ta}) within a radius of 2\arcmin--3\arcmin.
We have measured the radial distribution of the elemental abundances of O, Ne, Mg, Si, S, Ar, Ca, and Ni as well as that of Fe.

We compared our abundances with those obtained with {\it ASCA} (\cite{dupke}) in Fig.~\ref{fig:r-abun}.
In general, the results from both satellites are consistent with each other within their 90\% confidence limits.
The larger effective area and better spatial resolution of {\it XMM-Newton} improved the accuracy of the measurements significantly.
For example, 
a central abundance increase of Si, which was suggested by the {\it ASCA} measurements (\cite{dupke}; Finoguenov et al. \cite{finoguenov00}), 
has now been detected  confidently with {\it XMM-Newton}.
Furthermore, the O abundance was determined accurately for the first time so that we could detect a significant contrast 
between the O and Si-S-Fe abundances profiles (or a change in O/Fe ratio over the cluster).
The Ne and Mg abundances are also statistically consistent with being constant over the cluster area.
In short, a significant difference in the radial abundance profiles of O-Ne-Mg as compared to Si-S-Fe-Ni has been observed for the first time in this cluster.

\subsection{Origin of the metals in the ICM}
The settling of heavy elements towards the cluster core due to diffusion 
takes longer than the age of the cluster (Sarazin \cite{sarazin}).
Therefore the varying abundance ratio within the cluster implies that 
there are at least two different origins for the metals in the ICM.
This argument is independent of any theoretical model for the metal production.

What is the origin of the difference in the abundance profile between O-Ne-Mg and Si-S-Fe-Ni ?
An increase of the relative contribution from SNe Ia to the metal production towards the center is one possibility.
This is because a SN Ia is supposed to produce O/Fe, Ne/Fe, Mg/Fe ratios smaller than the solar value.
Then what causes the central increase in the Fe abundance in terms of an enhancement contribution from SNe Ia?
The metal-rich gas produced in the cD galaxy is the most natural origin of the Fe increase as demonstrated by a comparison of clusters with and without a cD galaxy (e.g. Makishima et al. \cite{makishima}; references therein; De Grandi and Molendi \cite{grandi}).

To compare the result obtained with the above idea quantitatively,
we presume that the excess of the metal abundances is solely produced by SN Ia in the cD galaxy over the galaxy life-time.
Firstly, we examine this idea in terms of the total Fe mass in the ICM.
Within the radius of 84~\arcsec\ from the cluster center
the integrated iron mass in the ICM is $3.2\times 10^{8}~h^{-2.5}$~\hbox{M$_{\odot}$} with our measurements of the ICM density and iron abundance.
Within the same area, 
the total blue luminosity of the cD galaxy is $4.1\times 10^{10}$~\hbox{L$_{\odot}$}$h^{-2}$ (Postman and Lauer \cite{postman}).
These give an excess ($\frac{1}{2}$ of the total mass) iron mass-to-light ratio around the cD galaxy of $4\times 10^{-3}h^{-0.5}$ in the solar units .
This is consistent with the value estimated by Renzini et al. (\cite{renzini}), 
who assumed the standard SN Ia rate in elliptical galaxies and 15~Gyr duration,
within an uncertainty of the SN Ia rate (a factor of $\sim 4$).
Therefore we conclude that the excess Fe mass in the ICM around the cD galaxy can be supplied by the standard SN Ia frequency over a Hubble time.
Fukazawa et al. (\cite{fukazawa98}) found that the excess Fe mass around the cD galaxy in other clusters can also be produced by the standard SN Ia rate over a Hubble time.
Secondly, we examined our results by comparing the abundance ratio of the excess metals with those based on the SN Ia model (Nomoto et al. \cite{nomoto97a}; Table~\ref{tbl:abun-ratio}).
To estimate the excess abundance, 
we fitted the obtained abundance profiles (Fig.~\ref{fig:r-abun}) with linear functions and
subtract the abundance in the outer-most bin (128\arcsec--384\arcsec\ in radius) from that of the central bin (0\arcsec--48\arcsec).
Relative to the large yield of Fe,
a SN Ia produces only a small amount of O, Ne, and Mg.
This is consistent with 'no excess' of O, Ne, and Mg in our observation.
Although the excess abundances of Si and S qualitatively agree with this idea, 
Si and S exhibit significantly ($>90$\% confidence limit) larger abundance than the prediction.
The excesses of Ar, Ca, and Ni are marginally consistent with the prediction.
Interstingly, relatively large Si/Fe and S/Fe are also suggested in the core of M87 (Molendi and Pizzolato \cite{molendi01b}).

In summary, the mass excess of Fe as well as Ar, Ca, and Ni and lack of excess of O, Ne, and Mg in the core 
are consistent with the assumption that the metal excess is solely produced by SNe Ia in the cD galaxy.
However, Si and S are not fully consistent with the idea.
This suggests an additional source for the metal excess.

Beyond the cD galaxy region, the abundance ratios with respect to Fe
are all consistent with the solar values except for S/Fe which is marginally smaller than the solar value (Table~\ref{tbl:abun-ratio}).
We compared the observed abundance ratios with the model prediction for SN Ia (Nomoto et al. ~\cite{nomoto97a}) and SN II (Nomoto et al. ~\cite{nomoto97b}; Arnett \cite{arnett}; Tsujimoto et al. \cite{tsujimoto}; Maeder \cite{maeder}; Woosley and Weaver \cite{woosley}).
For the SN II model, 
we used calculations by \cite{dupke} and Gibson et al. (\cite{gibson}) who averaged elemental yields over the progenitor mass range 10--50~\hbox{M$_{\odot}$} for a Salpeter IMF.
As illustrated in Table~\ref{tbl:abun-ratio}, the SN II abundances have uncertainties of a factor $\sim 2$ depending on the adopted model.

The observed abundance pattern in the outer part of the cluster cannot be solely produced by SN Ia. 
For example, the O, Ne, and Mg abundance ratios are at least 10 times larger than the SN Ia prediction.
At the same time, this pattern is inconsistent with the SN II prediction, 
even if we take into account the relatively large uncertainties in the model.
In fact, 
all the observed abundance ratios are between the two predictions, 
except for Ar which has however a large uncertainty.
Consequently, we conclude that the metals in the outer region of the ICM were produced by a mix of both types of SNe.
The relative fraction of each type of SN can be estimated from the 
observed abundance ratios of any two elements and the model yields of these two elements.
However, 
due to the large uncertainties in the yields for the SN II model, 
the results are strongly SN II model dependent.
For example, 
the observed values of O/Fe and Si/Fe 
imply an Fe mass fraction due to SN Ia of 0.4--0.7 and 0.7--0.8, respectively.
Here we used the range of SN II yields calculated by Gibson et al. (\cite{gibson}) which correspond to the last row of Table~\ref{tbl:abun-ratio}.

\begin{table*}
\caption[]{Abundance ratios and their 68\% confidence limits along with theoretical model prediction from SN Ia and SN II.
The abundances are relative to Fe normalized to the solar value, i.e (X/Fe)/(X$_{\odot}$/Fe$_{\odot}$) where X and Fe are number density of the element and Fe.}
\label{tbl:abun-ratio}
\begin{center}
\begin{tabular}{lcccccccc}
\hline 
	& O		& Ne	& Mg	& Si	& S	& Ar	& Ca	& Ni \\
\hline \hline
Excess$^a$	& $-0.02\pm0.37$  & $0.20\pm0.61$  & $0.05\pm0.54$  & $1.31\pm0.24$  & $1.45\pm0.34$  & $1.55\pm0.78$  & $0.34\pm0.92$ & $2.57\pm1.18$ \\
Cluster$^b$ & $0.87\pm0.28$ & $1.19\pm0.38$ & $1.03\pm0.41$ & $1.26\pm0.20$ & $0.58\pm0.24$ & $0.39\pm0.59$ & $1.45\pm0.64$ & $1.84\pm0.91$ \\
\hline
SN Ia$^c$& 0.035	& 0.006	& 0.035	& 0.5	& 0.5	& 0.5	& 0.5	& 4.8\\
SN II$^c$& 3.7		& 2.5	& 3.7	& 3.7	& 2.5	& 1.7	& 1.7	& 1.7\\	
SN II$^d$& 1.6--3.3	& 1.4--2.8 & 1.4--2.8 & 2.7--3.8 & 1.7--4.1 & - & - & - \\
\hline
\end{tabular}
\end{center}
\begin{description}
\item[$^a$] The difference in abundance between the central and outer regions.
\item[$^b$] The observed abundance ratio in 128\arcsec--384\arcsec\ in radii from the cluster center.
\item[$^c$] Nomoto et al. (\cite{nomoto97a}) and Nomoto et al.(~\cite{nomoto97b}) after \cite{dupke}.
\item[$^d$] Gibson et al. (\cite{gibson}) who employed several SN II model yields.
\end{description}
\end{table*}

\subsection{Resonant scattering}
As pointed out by Gil'fanov et al. (\cite{gilfanov}), 
resonant scattering could affect the abundance measurements at some cluster cores.
In the temperature range of A~496, 2--3 keV, 
significant resonant scattering is expected for the \ion{Fe}{xxv} (1.9~\AA), \ion{Fe}{xxiv} (10.6~\AA) and \ion{Fe}{xxiii} (11.0~\AA) lines.
The optical depth, $\tau$, of these lines in A~496 towards the cluster center
can be estimated to be 3--4, assuming $h=0.7$.
In this case, the equivalent width can be depressed by 50\% at most within the cluster core (Gil'fanov et al. \cite{gilfanov}).
Therefore, the profiles of the Fe abundance and hence the temperature within the core ($\sim 30$~\arcsec\ in radius) shown in Fig.~\ref{fig:r-ta} potentially may have been distorted.

For the RGS spectral fitting, 
we integrated all emission within a full-width of 1.1\arcmin.
Furthermore, the other Fe lines detected in the RGS spectra which dominate the line emission (\ion{Fe}{xxiii} at 11.74 and 12.19~\AA, \ion{Fe}{xxiv} at 11--15~\AA\ in the rest frame) 
cannot be affected by resonant scattering since these transitions do not end at the ground-state of the ion.
Therefore, the net effect on the RGS result (Table~\ref{tbl:rgs-fits}) should be much smaller than a 50\% depression.
To examine this further, 
we fitted the central RGS spectra excluding the Fe line wavelength band (10.4~\AA--11.2~\AA\ in the rest frame), where potentially optically-thin lines are found.
We confirmed that the derived parameters do not really change within the errors. 

Other lines should have much smaller $\tau$ than 3--4.
For example, the \ion{O}{viii} and \ion{Si}{xiv} ${\rm Ly\alpha}$ lines, 
which mainly determine the abundance of O and Si,
have $\tau$ smaller than that of the \ion{Fe}{xxiv} line by a factor of 2--3.
In this case, the depression of the equivalent width and hence the derived abundance
is 30\% at most for the integrated spectrum within the core, which corresponds to the central bin in Fig.~\ref{fig:r-abun}.
Therefore we can rule out that resonant scattering is the cause of the difference in the radial profile abundance between O-Ne-Mg and Si-S-Fe-Ni.

\subsection{Central Cool Component}
The calculated radiative cooling time is below a Hubble time within the central radius of 2\arcmin--3\arcmin\
as already pointed out e.g. by Nulsen et al. (\cite{nulsen}).
At the cluster center, the cooling time is less than $10^9$ yr.
Therefore we expect plasma with a range of temperatures less than that of the ambient ICM ($\sim 4$~keV), 
unless some heat source effectively balances the cooling.
Nevertheless, the temperature profile derived from spatially resolved spectra (Fig.~\ref{fig:r-ta})
indicates that the ICM cools down only to 1.8~keV.
Consistently, the RGS spectra indicate that 
the mass flow rate should be small ($<10$~$h^{-2}$~\hbox{M$_{\odot}$}yr$^{-1}$) when an isobaric cooling flow was assumed.
Alternatively, the emission should have a cut-off at $\sim 1$~keV for a moderate level of the cooling flow.
There is no spectral evidence for gas cooler than 1.0~keV.
The absence of emission from cool gas has been found also 
in A~1835 (Peterson et al. \cite{peterson}), S\'ersic 159-03 (Kaastra et al. \cite{kaastra01}), and A~1795 (Tamura et al. \cite{tamura}),
indicating that this emission property is general for clusters with a cD galaxy.
Peterson et al. (\cite{peterson}), Fabian et al. (\cite{fabian01}), and Molendi and Pizzolato (\cite{molendi01}) discussed  several possibilities to explain these observations.

\begin{acknowledgements}
This work is based on observations obtained with {\it XMM-Newton}, an ESA science 
mission with instruments and contributions directly funded by 
ESA Member States and the USA (NASA).
The SRON-institute for Space Research Utrecht is supported
financially by NWO, the Netherlands Organization for Scientific
Research. 
We thank J.W. den Herder and anonymous referee for useful comments.
\end{acknowledgements}

\end{document}